\documentclass[onecolumn,superscriptaddress,secnumarabic,amssymb,amsmath,nobibnotes,aps,prd,nofootinbib,altaffilletter]{revtex4}
\usepackage{amsmath,amssymb}
\usepackage{graphicx}
\usepackage{dcolumn}
\usepackage{bm,color}
\usepackage{hyperref}
\usepackage{accents}
\usepackage{amssymb,float}
\usepackage{amsmath}
\usepackage{multirow}
\usepackage{siunitx}
\usepackage{tabularx}
\usepackage{booktabs}
\usepackage{url}
\usepackage{orcidlink}

\usepackage{capt-of} 
\usepackage{placeins} 
\usepackage{psfrag}
\usepackage{times}
\usepackage[varg]{txfonts}
\usepackage{xcolor}
\usepackage{xspace}

\begin{document}

\title{Phantom scalar field cosmologies constrained by early cosmic measurements}

\author{José Antonio Nájera\orcidlink{0000-0001-9738-7704}}
\email{joseantoniodejesus.najeraquintana@studenti.unipd.it}
\affiliation{ Dipartimento di Fisica e Astronomia “Galileo Galilei”, Università degli Studi di Padova \\
via Marzolo 8, I-35131, Padova, Italia}

\author{Celia Escamilla-Rivera\orcidlink{0000-0002-8929-250X}}
\email{celia.escamilla@nucleares.unam.mx}
\affiliation{Instituto de Ciencias Nucleares, Universidad Nacional Aut\'{o}noma de M\'{e}xico, 
Circuito Exterior C.U., A.P. 70-543, M\'exico D.F. 04510, M\'{e}xico.}


\begin{abstract}
In this work, we explore new constraints on phantom scalar field cosmologies with a scalar field employing early times catalogues related to CMB measurements, along with the local standard observables, like Supernovae Type Ia (SNIa), $H(z)$ measurements (Cosmic Clocks), and Baryon Acoustic Oscillations (BAO) baselines. In particular, we studied a tracker phantom field with hyperbolic polar coordinates that have been proposed in the literature. The main goal is to obtain precise cosmological constraints for $H_0$ and $\sigma_8$, in comparison to other constructions that present tension in early cosmological parameters. Our results show that phantom scalar field cosmologies have a reduced statistical tension on $H_0$ that it is less than 3$\sigma$ using model-independent CMB catalogues as SPT-3G+WMAP9 and ACTPol DR-4+WMAP9 baselines. This suggests these models using a different phantom potential might address the Hubble constant problem and reduce the systematics involved. 
\end{abstract}

\maketitle

\section{Introduction}

One of the first observational evidence of the late-time cosmic acceleration was shown through measurements of Type Ia Supernovae (SNIa) \cite{SupernovaCosmologyProject:1998vns, SupernovaSearchTeam:1998fmf}. 
Subsequently, more observables confirmed these results, e.g. the Cosmic Microwave Background (CMB) \cite{Boomerang:2000jdg}, Baryon Acoustic Oscillations (BAO) \cite{SDSS:2005xqv, BOSS:2013rlg}, and weak gravitational lensing \cite{DES:2017myr}. 
Moreover, the capability of SNIa's to prove the cosmic acceleration is based on the facts that these objects are bright enough to be seen at large redshift and in large quantities with a precision of $\sim 0.1$ mag in brightness \cite{Brout:2022vxf}. However, these observables have considerably been minimised by their associated statistical errors and the uncertainties in estimating cosmological parameters \cite{Scolnic:2019apa, Lu:2022utg}. 

Furthermore, through the CMB temperature and polarization anisotropy observations we have been capable of setting a strong confirmation of the standard so-called $\Lambda$-Cold Dark Matter ($\Lambda$CDM) model associated with the structure formation. This model has the advantage of explaining with quite precision of other observations at different redshift ranges and also explaining the effects of the cosmic acceleration through the so-called \emph{dark energy} (DE). However, with the increase in surveys and their experimental sensitivity, 
significant cosmological statistical tensions have been raised and related to certain discrepancies that could be due to systematic errors or modifications (or extensions) to the standard $\Lambda$CDM model \textit{per se}. While the persisting tensions of the CMB bring to discuss issues associated with $S_8$ with cosmic shear data \cite{DiValentino:2020vvd}, $\Omega_k$ different from zero (flat cosmology) \cite{DiValentino:2019qzk}  and $A_L$ internal anomaly \cite{DiValentino:2020srs}, the Hubble constant with local measurements, $H_0$ \cite{DiValentino:2020zio} is the most statistically significant tension at more than $\sim$4$\sigma$. In such case, CMB cosmological constraints, in specific, Planck constraints \cite{Planck:2018vyg}, can be obtained in a model-dependent way through the assumption of the $\Lambda$CDM model, i.e. if we change the assumption, the constraints will change. These tensions have reached a level of statistical significance that understanding their physics is of importance for precision cosmology. However, if these issues are not related to systematics, they could represent a crisis for the $\Lambda$CDM model, and their observational confirmation could bring a change in our current conceptions of the evolution and structure of the universe. Along with these ideas, several proposals from theoretical and systematic points of view have been developed in the last years \cite{Abdalla:2022yfr}, and references therein. However, none of them have reached a fully solved foreground yet. This guides us to explore other schemes that could shed some light on the cosmological tensions, in particular on $H_0$ and $\sigma_8$ tensions.

From the surveys point of view, at local scales, several missions have been working to find better cosmological constraints adjoint with better systematics and increasing data baselines. Some of them as large-scale structure (LSS) observations with measurements from the Dark Energy Survey (DES) \cite{DES:2016jjg}, the Dark Energy Spectroscopic Instrument (DESI) \cite{DESI:2016fyo}, the Legacy Survey of Space and Time (LSST) on the Vera Rubin Observatory \cite{LSSTDarkEnergyScience:2018jkl}, and Euclid \cite{Amendola:2016saw}, among others, have extended the concordance cosmological model to include EoS parameters of dark energy with some shifts within 1$\sigma$. At early scales, missions such as Planck 2018 \cite{aghanim2020planck}, ACTPol DR-4 \cite{aiola2020atacama}, and SPT-3G \cite{dutcher2021measurements}, have been working with CMB polarisations regarding in the likelihood precision measurements. 

On this line of thought, an interesting path to study both the nature of dark energy and relax (even solve) cosmological tensions, has been to consider directly: \textit{(i)} a negative Equation--of--State (EoS) constant value \cite{Escamilla-Rivera:2019aol, Escamilla-Rivera:2016qwv,Escamilla:2023oce}. \textit{(ii)} A dynamical EoS \cite{Escamilla-Rivera:2021boq, Escamilla-Rivera:2019hqt,Zhang:2023zbj}. \textit{(iii)} A dark energy component associated with extra terms obtained from first principles in alternative theories of gravity \cite{Jaime:2018ftn, Clifton:2011jh,Amendola:2006we} and, \textit{(iv)} a dark energy foreground in extended theories of gravity \cite{Bahamonde:2021gfp} and references therein, which shows a good promise to obtain a late cosmic acceleration. On the second classification, scalar fields with kinetic terms and potentials are considered as candidates to explain the dynamics of dark energy, e.g quintessence \cite{deRitis:1999dt,Alho:2023xel}, $K$-essence, etc. However, these examples suffer from theoretical issues where values of $w_{DE}<-1$ are not allowed, this is the so-called phantom limit. To achieve a viable evolution for dark energy, we can propose \textit{quintom models}~\cite{cai2010quintom,setare2008coupled}, which denote the dynamics from quintessence scalar fields plus a phantom phase. This feature could be a good landscape to address the $H_0$ tension.
Some effort on this scheme has proposed a statistical analysis of an inverse power law quintessence model constrained by Dark Energy Task Force simulated baseline \cite{Yashar:2008ju}, cosmological constraints on quintom tracker solutions using SNIa, BAO and the compressed Planck likelihood \cite{Urena-Lopez:2020npg,Roy:2023vxk,Adil:2022hkj}, which brings high correlations with the $\Lambda$CDM model due the early times baselines calibrated with it. Some successful scenarios include a large class of quintessence models that experiment with an early dark energy phase \cite{Copeland:2023zqz}, a model that has been proposed as a solution to the Hubble tension.

Since these approaches include early time baselines that are calibrated with the standard cosmological model, $\Lambda$CDM, in this work we incorporate three CMB baselines that are model-independent and allow us to obtain in phantom scalar field cosmologies a reduced statistical tension on $H_0$
less is than 3$\sigma$ for the SPT3G+WMAP9 and ACTPol+WMAP9 baselines.

This work is divided as follows:
In Sec.~\ref{sec:quintessenceback} we summarise the reconstruction of tracker quintessence cosmologies background and the most promising scenarios available to relax the $H_0$ tension. All of these cases are described through their Friedmann evolution equations. Furthermore, we are going to consider the standard $\Lambda$CDM model in addition to these reconstructions.
In Sec.~\ref{sec:method} we present the statistical methodology employed for the baseline datasets mentioned. We divided our analysis into baseline (low-z) local observations and a high-z observables.
Our results on new constraints and cosmological tensions discussions are presented in Sec.~\ref{sec:results}.
Finally, our conclusions are given in Sec.~\ref{sec:final}.


\section{Phantom Scalar Field cosmology background}
\label{sec:quintessenceback}

Phantom scalar field dark energy models are described by an EoS that satisfies $w<-1$ \cite{caldwell2002phantom}. Even though the current observational data have shown a good agreement with the standard cosmological model plus a cosmological constant, some of these observational baselines suggest the presence of a phantom divide boundary, i.e. dynamical dark energy proposals. This aspect could be a signature of dynamical dark energy that can be well constrained by observational data, however, denotes a big challenge in precision cosmology. Furthermore, these models have introduced scalar fields to explore the dynamics of evolving dark energy through this phantom divided line, having issues like fine-tuning. Within this scheme, tracker scenarios have been considered \cite{ratra1988cosmological, steinhardt1999cosmological} where the scalar field controls the energy density and reproduces attractor background solutions, which can be constrained with local observations. In this work, we will study the parameterisations proposed in \cite{urena2020generalized,roy2018new} and in particular, the phantom scalar field dark energy model proposed in \cite{cedeno2021tracker}. Considering several CMB baselines, we will verify if this model can solve (or relax) the $H_0$  tension. We will employ model-independent CMB experiments, and for comparison, we include Planck 2018 constraints \cite{aghanim2020planck} since this baseline is calibrated with the standard $\Lambda$CDM model. 

To incorporate a phantom scalar field term, we start with a gravitational action given by
\begin{equation}\label{eq:action}
    S = \frac{1}{16\pi G} \int d^4x \sqrt{-g} \left( R + \mathcal{L}_\phi (\phi, \partial_\mu \phi) + \mathcal{L}_m \right),
\end{equation}
where $R$ is the Ricci scalar, $\mathcal{L}_\phi$ is the scalar field Lagrangian (for a phantom scalar field, it is given by $\mathcal{L}_\phi = -1/2 \partial^\mu \phi \partial_\mu \phi - V(\phi)$ where $V(\phi)$ is the potential of the scalar field) and $\mathcal{L}_m$ is the matter Lagrangian. We will work in a flat Friedman-Lemaître-Robertson-Walker (FLRW) metric. By varying the action (\ref{eq:action}) with respect to the FLRW metric and to the scalar field $\phi$ and setting it to zero, we obtain the following Friedmann equations 
\begin{eqnarray}
    H^2 &=& \frac{8\pi G}{3} \left( \rho + \rho_\phi \right), \\
    \dot{H} + H^2 &=& -\frac{4\pi G}{3} (\rho + 3p + \rho_\phi + 3p_\phi), \\
    \dot{\rho} + 3H(\rho + p) &=& 0,
\end{eqnarray}
and the Klein-Gordon equation for the scalar field given by
\begin{equation}
\label{eqn:Klein-Gordon}
    \Ddot{\phi} +3H\dot{\phi} + \frac{\partial V}{\partial \phi} = 0,
\end{equation}
where $\rho$ and $p$ are the density and pressure of the matter species, $\rho_\phi$ and $p_\phi$ the density and pressure of the scalar field $\phi$. The density and pressure of the phantom scalar field are given by
\begin{equation}
\label{eqn:scalarFieldDensity}
    \rho_\phi = -\frac{1}{2} \dot{\phi}^2 + V(\phi), \quad \quad p_\phi = -\frac{1}{2} \dot{\phi}^2 - V(\phi),
\end{equation}
and as we can see, these fields give rise to an equation of state $w<-1$. In this framework, it is customary to consider a new set of hyperbolic polar coordinates which ease the numerical computations \cite{cedeno2021tracker}
\begin{eqnarray}\label{eq:sys}
    x &=& \sqrt{\dfrac{8\pi G}{6}}\frac{\dot{\phi}}{H} = \sqrt{\Omega_\phi} \sinh (\theta/2), \quad \quad
    y = \sqrt{\dfrac{8\pi G V}{3}}\frac{1}{H} = \sqrt{\Omega_\phi} \cosh(\theta/2),\\
    y_1 &=& -2\frac{\sqrt{2}}{H} \frac{\partial \sqrt{V}}{\partial \phi}, \quad \quad \quad \quad
    y_2 = -\frac{4}{H}\sqrt{\frac{3}{8\pi G}} \frac{\partial^2 \sqrt{V}}{\partial \phi^2},
\end{eqnarray}
which enables us to rewrite the Klein-Gordon equation (\ref{eqn:Klein-Gordon}) as
\begin{eqnarray}
    \Omega'_\phi &=& 3aH \Omega_\phi [w_{\text{tot}} + \cosh(\theta_\phi], \\
    \theta' &=& -aH\left[3\sinh(\theta) + y_1\right],\\
    y'_1 &=& aH\left[ \frac{3}{2} (1+w_{\text{tot}}) y_1 + y_2(\Omega_\phi, \theta, y_1) \right],
\end{eqnarray}
where the prime $'$ denotes the derivative with respect to conformal time. Also, $w_{\text{tot}} = p_{\text{tot}}/\rho_{\text{tot}}$, is the total EoS including the matter species and the scalar field density and pressure. Notice that in this set of variables we need to specify the function $y_2$ which is equivalent to specifying the potential $V(\phi)$. Moreover, by using the definitions of scalar field density and pressure (\ref{eqn:scalarFieldDensity}), we obtain
\begin{eqnarray}
    \rho_\phi = \frac{3H^2 \Omega_\phi}{8\pi G}, \quad \quad
    p_\phi = -\rho \cosh(\theta).
\end{eqnarray}
Then, $\Omega_\phi$ is the density parameter for the phantom scalar field. On the other hand, the EoS of this field is given by $w_\phi = -\cosh(\theta)$. The remaining part to solve the differential equations is to determine $y_2$. A parameterisation can be considered \cite{urena2020generalized,roy2018new}
\begin{equation}
\label{eqn:y_2}
    y_2 = y\left(\alpha_0 + \alpha_1 \frac{y_1}{y} + \alpha_2 \frac{y^2_1}{y^2} \right),
\end{equation}
where $y$ is given by Eq.~(\ref{eq:sys}) and $\alpha_0, \alpha_1, \alpha_2$ are free parameters of the phantom field. This form of $y_2$ allows us to consider a wide variety of potentials like thawing and freezing potentials \cite{roy2018new} depending on the values of the free parameters $\alpha_0, \alpha_1, \alpha_2$. 

However, even with this proposal, it was found that this kind of phantom model does not address the Hubble constant problem when considering a compressed Planck likelihood. This analysis gave a Hubble value of $H_0 = 69.1^{+0.5}_{-0.6}$ km/s/Mpc \cite{cedeno2021tracker}, which is in a $3\sigma$ C.L. tension with the latest result from the SH0ES collaboration \cite{riess2022comprehensive}. In this work, we will reconstruct the constraints using the full Planck \cite{aghanim2020planck} likelihood. Furthermore, we will compare this constraint with the ones using model-independent CMB baselines. Within the baselines under consideration we have the ACTPol DR-4 \cite{aiola2020atacama}, the SPT-3G \cite{dutcher2021measurements} and the WMAP9 \cite{hinshaw2013nine} datasets. This will determine whether a phantom scalar field model like this can predict a larger late-time expansion of the universe than $\Lambda$CDM.


\section{Methodology}
\label{sec:method}

To solve the above equations, we used our modified version of the Boltzmann code \texttt{CLASS} \footnote{\url{https://lesgourg.github.io/class_public/class.html}} \cite{blas2011cosmic}, interfaced to the sampling code \texttt{MontePython} \footnote{\href{https://github.com/brinckmann/montepython_public}{brinckmann/montepython\_public}} \cite{audren2013conservative,brinckmann2019montepython}. One common core of the code is the initial conditions to perform the integrals, which were suggested in \cite{cedeno2021tracker}. Following this suggestion, we considered the following initial conditions for $\theta, y_1, \Omega_\phi$ 
\begin{eqnarray}
    \cosh(\theta_i) = 1 + \frac{2}{3\alpha_2}, \quad\quad
    y_{1i} = -3\sinh(\theta_i), \quad\quad
    \Omega_{\phi i} = A \Omega_{\phi 0} a_i^{4(1+1/(2\alpha_2))} \left( \frac{\Omega_{m0}}{\Omega_{r0}} \right)^{1+1/(2\alpha_2)},
\end{eqnarray}
where the sub-index 0 means at present time. $\Omega_{m0}, \Omega_{r0}, \Omega_{\phi 0}$ are the matter, radiation and phantom scalar field density parameters at the current time. $a_i$ is the scale factor at which \texttt{CLASS} starts the integration. We select $a_i = 10^{-14}$ by default. 
However, an important difference in our numerical code adaption is that we did not include the presence of a cosmological constant $\Lambda$. Thus, we are letting the phantom scalar field drive the cosmic late-time accelerated expansion. Furthermore, $A$ is a tuning parameter which varies until we satisfy the closure relation $\Omega_{r0} + \Omega_{m0} + \Omega_{\phi 0} = 1$. We are considering a universe filled with matter, radiation and a phantom scalar field. 

\subsection{Observational baselines}

Our study is based on constraining the full tracker phantom model using two kinds of baseline: late-time and early-time surveys. According to this division, we can consider the following:
\begin{itemize}
\item Late-time baselines:
\begin{enumerate}
       \item \textbf{Supernovae Type Ia (SNIa) Pantheon:} We use the 1048 data points provided by the \textit{Pantheon}       \cite{scolnic2018complete}. This baseline
        measure the apparent distance for several SNIa events in $0.01 < z < 2.3$. Furthermore, this catalog provides SN magnitudes corrected for the stretch and colour effects along with the maximum brightness, the mass of the host galaxy, and sky position bias. To compute a cosmological useful quantity we can calculate the distance modulus $\mu = m - M$, where $M$ is the absolute magnitude that is considered a fixed value for our analyses. Furthermore, the $\chi^2_{\text{SN}}$  for the Pantheon sample is
\begin{equation}
    \chi^2_{\text{SN}} = \Delta\mu(z_i,\Theta)^TC^{-1}_{\text{SN}} \Delta \mu(z_i,\Theta) + \ln\left(\frac{S}{2\pi}\right) - \frac{k^2(\Theta)}{S},
\end{equation}
where $C^{-1}_{\text{SN}}$ is the total covariance matrix for the data, $S$ is the sum of all components of the inverse of the matrix and $k(\Theta) = \Delta\mu(z_i,\Theta)^TC^{-1}_{\text{SN}} $, using $\Delta\mu(z_i,\Theta) = \mu(z_i,\Theta) - \mu_{\text{obs}}(z_i)$. Also, the distance modulus $\mu(z)$ can be compute using the expression:
\begin{equation}
    \mu(z_i,\Theta) = 5\log[D_L(z_i,\Theta)] + M, 
\end{equation}
and where $D_L(z_i,\Theta)$ is the luminosity distance given as: 
\begin{equation}
    D_L(z_i,\Theta) = c(1+z_i)\int_0^{z_i} \frac{dz'}{H(z',\Theta)},
\end{equation}
where $c$ is the speed of light and $H(z_i,\Theta)$ is the Hubble parameter. 
           \item \textbf{Cosmic clocks (CC):} 
         This sample offers a good tool to constrain the Hubble rate $H(z)$ at different $z$. To this end, the final catalog considered comes from the differential age method \cite{jimenez2002constraining}. In particular, we consider the cosmic clocks 2016 catalog \cite{moresco20166}.  The CC method consists in using spectroscopic dating techniques on passively-evolving galaxies to compute the age difference between two galaxies at different $z$. By measuring this age difference, $\Delta z / \Delta t$ we can be compute $H(z) = -(1+z)^{-1} \Delta z/ \Delta t$. 
For our MCMC analysis, we compute $\chi^2_\mathrm{CC}$ to compare the agreement between the theoretical Hubble parameter values $H(z_i,\Theta)$, with model parameters $\Theta$, and the observational Hubble data values $H_{\mathrm{obs}}(z_i)$, with an observational error of $\sigma_H(z_i)$. Therefore, the $\chi^2_\mathrm{CC}$ is calculated using the following expression:
\begin{equation}
    \chi^2_\mathrm{CC} = \sum^{31}_{i=1} \frac{\left(H(z_i,\Theta) - H_{\mathrm{obs}}(z_i)\right)^2}{\sigma^2_H(z_i)} \,.
\end{equation}

    \item \textbf{Baryon Acoustic Oscillations (BAO):} 
    In this work, we include measurements of the Hubble parameter and the corresponding comoving angular diameter at $z_{\mathrm{eff}} = {0.38,0.51}$, which were obtained from the third generation of the SDSS mission (SDSS BOSS DR12)  \cite{alam2017clustering}.
For this BAO baseline we compute the Hubble distance $D_H(z)$ given by $D_H(z) = \frac{c}{H(z)}$. We also use the angular diameter distance $D_A(z)$ given by
\begin{equation}
    D_A(z) = \frac{c}{1+z} \int^z_0 \frac{dz'}{H(z')} \,,
\end{equation}
where the first is the comoving angular diameter distance $D_M$ given trough $D_M = (1+z)D_A(z)$, and the second one is the volume average-distance given by
\begin{equation}
    D_V(z) = (1+z)^2 \left[D_A(z)^2 \frac{cz}{H(z)}\right]^{\frac{1}{3}} \,. 
\end{equation}
Afterwards, we calculate the corresponding combination of results $\mathcal{G}(z_i)=D_V(z_i)/r_s(z_d),\allowbreak\,r_s(z_d)/D_V(z_i),\allowbreak\,D_H(z_i),\allowbreak\,D_M(z_i)(r_{s,\mathrm{fid}}(z_d)/r_s(z_d)),\allowbreak\,H(z_i)(r_s(z_d)/r_{s,\mathrm{fid}}(z_d)),\allowbreak\,D_A(z_i)(r_{s,\mathrm{fid}}(z_d)/r_s(z_d))$. For this, we require the comoving sound horizon at the end of the baryon drag epoch at $z_d \sim 1059.94$ \cite{Planck:2018vyg} which can be calculated through
\begin{equation}
    r_s(z)=\int_z^\infty\frac{c_s(\tilde{z})}{H(\tilde{z})}\,\mathrm{d}z=\frac{1}{\sqrt{3}}\int_0^{1/(1+z)}\frac{\mathrm{d}a}{a^2H(a)\sqrt{1+\left[3\Omega_{b,0}/(4\Omega_{\gamma,0})\right]a}}\,,
\end{equation}
where we have considered a fiducial value of $r_{s,\mathrm{fid}}(z_d)=147.78,\mathrm{Mpc}$ \cite{Planck:2018vyg} with an assumption of $\Omega_{b,0}=0.02242$ \cite{Planck:2018vyg} and $T_{0}=2.7255$. The corresponding $\chi^2$ is given by
\begin{equation}
    \chi^2_{\text{BAO}}(\Theta) = \Delta G(z_i,\Theta)^T C_{\text{BAO}}^{-1}\Delta G(z_i,\Theta)
\end{equation}
where $\Delta G(z_i,\Theta) = G(z_i,\Theta)-G_{\text{obs}}(z_i)$and $C_{\text{BAO}}$ is the corresponding covariance matrix for the BAO observations. 
    
\end{enumerate}

\item Early-time baselines:
\begin{enumerate}
    \item \textbf{Planck 2018:} For this CMB observations, we took the high-$\ell$ TTTEE, low-$\ell$ EE, low-$\ell$ TT, and lensing likelihoods \cite{aghanim2020planck}. Furthermore, polarization and temperature TT-TE-EE baselines were used at high multipole likelihood \texttt{Plik} $30<\ell<2500$ and at low multipoles TT-EE for $0<\ell<30$.
    \item \textbf{SPT-3G:} We also used the CMB polarization observations from the SPT-3G detector \cite{dutcher2021measurements}. To use this data, we took the \texttt{MontePython} likelihood for SPT-3G devised in \cite{chudaykin2022exploring}.
    \item \textbf{ACTPol DR-4:} This is the third CMB catalog considered coming from the Data Release 4 measured by the Atacama Telescope (ACT) Collaboration \cite{aiola2020atacama}. To use this catalog along with \texttt{MontePython}, we utilised the \texttt{pyactlike} python package devised by the ACT Collaboration \footnote{\href{https://github.com/ACTCollaboration/pyactlike}{ACTCollaboration/pyactlike}}. This likelihood also includes a Gaussian prior on $\tau = 0.06 \, \pm \, 0.01$.
    \item \textbf{WMAP9:} The final CMB catalog was the Wilkinson Microwave Anisotropy Probe and we took the results from the ninth year \cite{hinshaw2013nine}. To use this catalog along with others in MontePython we used the \texttt{clik} software \footnote{\href{https://github.com/benabed/clik}{benabed/clik}} that enabled us to install the WMAP9 likelihood and use it inside MontePython. 
    \end{enumerate}
\end{itemize}

To sample the cosmological parameters of our tracker models, we consider three different baselines for the catalogues described:
\begin{enumerate}
    \item Planck 2018+BAO+Pantheon+Cosmic clocks,
    \item SPT-3G+WMAP9+BAO+Pantheon+Cosmic clocks,
    \item ACTPol DR-4+WMAP9+BAO+Pantheon+Cosmic clocks.
\end{enumerate}
As we can notice, we included BAO, Pantheon and Cosmic clocks in our three combinations. In addition to this, we varied the CMB catalogues. Planck data was included in the first combination but not in the other ones to have two Planck-independent combinations to see if this improves or changes the convergence of the scalar field and cosmological free parameters with particular emphasis on the Hubble parameter at $z=0$, $H_0$. 

\begin{table}[h]
\centering
\renewcommand{\arraystretch}{1.5}
\resizebox{0.7\textwidth}{!}{\begin{tabular}{|l @{\hspace{0.4 cm}} |c|c|c|c|c|c|c|c|c|c}
 \hline 
			\textbf{Parameter}	& \textbf{Planck 2018+Late}	& \textbf{SPT-3G+WMAP9+Late}     & \textbf{ACTPol DR-4+WMAP9+Late}\\
			 \hline\hline 
   $\omega_{cdm}$ & $0.1196^{+0.00097}_{-0.00096}$ & $0.1179^{+0.0015}_{-0.0016}$ & $0.1208 \pm 0.0015$ \\ \hline
   $100\omega_b$ & $2.241 \pm 0.014$ & $2.265\pm 0.022$ & $2.239 \pm 0.0019$ \\ \hline
   $100\theta_s$ & $1.042^{+0.00028}_{-0.00029}$ & $1.04^{+0.00063}_{-0.00065}$ & $1.043^{+0.00059}_{-0.0006}$ \\\hline
   $n_s$ & $0.9668 \pm 0.0038$ & $0.9741^{+0.0065}_{-0.0067}$ & $0.9731^{+0.0046}_{-0.0045}$ \\\hline
   $\tau_{reio}$ & $0.05506^{+0.007}_{-0.0075}$ & $0.07805^{+0.011}_{-0.012}$ & $0.06893^{+0.0072}_{-0.0075}$ \\ \hline\hline
   $\alpha_0$ & $1.083^{+11}_{-3.9}$ & $5.567^{+6.4}_{-1.6}$ & $5.462^{+6.5}_{-1.6}$ \\ \hline
   $\alpha_1$ & $1.04^{+7}_{-2.5}$ & $4.603^{+3.4}_{-0.77}$ & $5.007^{+3}_{-0.71}$ \\ \hline
   $\alpha_2$ & $11.29^{+4.7}_{-1.5}$ & $4.061^{+0.79}_{-0.85}$ & $3.147^{+0.6}_{-0.53}$ \\ \hline\hline
   $z_{reio}$ & $7.731 \pm 0.73$ & $9.809^{+1.1}_{-1}$ & $9.114^{+0.7}_{-0.68}$ \\ \hline
   $YHe$ & $0.2479 \pm (5.9 \, \times 10^{-5})$ & $0.248\pm (9.3 \, \times 10^{-5})$ & $0.2478^{+8.1 \, \times 10^{-5}}_{-7.9 \, \times 10^{-5}}$ \\ \hline
   $H_0$ & $68.71^{+0.48}_{-0.57}$ & $70.01^{+0.55}_{-0.6}$ & $69.87^{+0.55}_{-0.61}$ \\ \hline
   $10^9 A_s$ & $2.103^{+0.029}_{-0.031}$ & $2.195^{+0.05}_{-0.055}$ & $2.185^{+0.032}_{-0.033}$ \\ \hline
   $\sigma_8$ & $0.815^{+0.0062}_{-0.0064}$ & $0.8291\pm 0.012$ & $0.8416^{+0.0093}_{-0.0096}$ \\ \hline
   $\Omega_m$ & $0.301 \pm 0.006$ & $0.2868^{+0.0069}_{-0.0072}$ & $0.2933^{+0.007}_{-0.0071}$ \\ \hline\hline
    $T_{H_0}$ & 4.01 & 2.80 & 2.92 \\ \hline
    \end{tabular}}
\caption{Mean values and uncertainties at $1\sigma$ C.L. for the phantom scalar field parameters. Notice that we included $10^9 A_s$ as a derived parameter instead of the traditional $\ln(10^{10} A_s)$. We present the results for our three baselines for comparison between different CMB catalogues: Planck 2018+BAO+Pantheon+Cosmic clocks, SPT-3G+WMAP9+BAO+Pantheon+Cosmic clocks, and ACTPol DR-4+WMAP9+BAO+Pantheon+Cosmic clocks. We abbreviate the late-time catalogues (BAO+Pantheon+Cosmic clocks) as Late. We also present the Hubble tension of each baseline in comparison to SH0ES 2022 \cite{riess2022comprehensive}. \label{tab:cosmologicalParametersPhantomScalarField}
}
\end{table}


\begin{table}[h]
\centering
\renewcommand{\arraystretch}{1.5}
\resizebox{0.7\textwidth}{!}{\begin{tabular}{|l @{\hspace{0.4 cm}} |c|c|c|c|c|c|c|c|c|c}
 \hline 
			\textbf{Parameter}	& \textbf{Planck 2018+Late}	& \textbf{SPT-3G+WMAP9+Late}     & \textbf{ACTPol DR-4+WMAP9+Late}\\
			\hline
   $\omega_{cdm}$ & $0.1191 \pm 0.00094$ & $0.1172 \pm 0.001$ & $0.1186^{+0.0013}_{-0.0014}$ \\ \hline
   $100\omega_b$ & $2.246\pm 0.014$ & $2.261^{+0.02}_{-0.021}$ & $2.243^{+0.018}_{-0.019}$ \\ \hline
   $100\theta_s$ & $1.042 \pm 0.00029$ & $1.041^{+0.00064}_{-0.00063}$ & $1.043^{+0.00061}_{-0.0006}$\\ \hline
   $n_s$ & $0.9681^{+0.0037}_{-0.0039}$ & $0.976^{+0.0062}_{-0.0063}$ & $0.9764^{+0.0044}_{-0.0045}$ \\ \hline
   $\tau_{reio}$ & $0.05707^{+0.0069}_{-0.008}$ & $0.08415 \pm 0.012$ & $0.07108^{+0.0075}_{-0.0076}$ \\\hline\hline
   $z_{reio}$ & $7.92^{+0.71}_{-0.76}$ & $10.36^{+1.1}_{-1}$ & $9.279^{+0.72}_{-0.69}$ \\ \hline
   $YHe$ & $0.2479 \pm (5.9 \, \times \, 10^{-5})$ & $0.2479^{+8.7\, \times\, 10^{-5}}_{-8.9 \, \times \, 10^{-5}}$ & $0.2479^{+7.8 \, \times \, 10^{-5}}_{-8 \, \times \, 10^{-5}}$ \\ \hline
   $H_0$ & $67.85^{+0.41}_{-0.43}$ & $68.17^{+0.4}_{-0.43}$ & $68.26^{+0.55}_{-0.56}$ \\ \hline
   $10^9 A_s$ & $2.109^{+0.029}_{-0.033}$ & $2.22^{+0.05}_{-0.056}$ & $2.183^{+0.033}_{-0.034}$ \\ \hline
   $\sigma_8$ & $0.8102^{+0.006}_{-0.0064}$ & $0.8254 \pm 0.011$ & $0.8263^{+0.0083}_{-0.0082}$ \\ \hline
   $\Omega_m$ & $0.3075^{+0.0057}_{-0.0056}$ & $0.301^{+0.0054}_{-0.0053}$ & $0.3028^{+0.0073}_{-0.0076}$ \\ \hline\hline
   $T_{H_{0}}$ & 4.87 & 4.60 & 4.28 \\ \hline
       \end{tabular}}
   \caption{Mean values and uncertainties at $1\sigma$ C.L. for the $\Lambda$CDM cosmological parameters. Notice that we include$10^9 A_s$ as a derived parameter instead of the traditional $\ln(10^{10} A_s)$. We present the results for the same baselines as for the phantom scalar field for comparison. We also present the Hubble tension of each baseline in comparison to SH0ES 2022 \cite{riess2022comprehensive}.\label{tab:cosmologicalParametersLCDM}
}
\end{table}


\begin{figure}[h]
\centering
\includegraphics[width=12.cm]{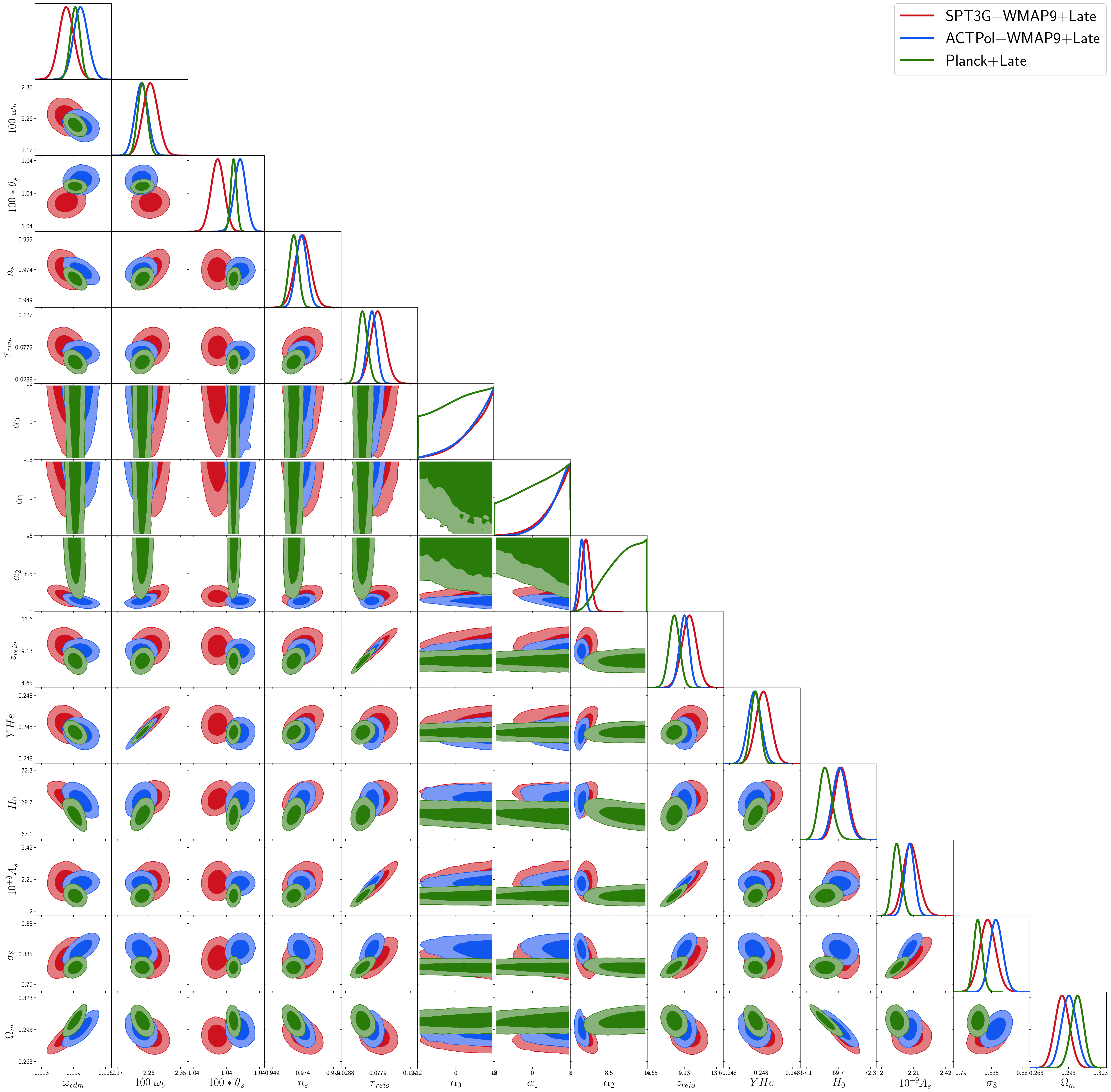}
\caption{$1 \sigma$ and $2\sigma $ confidence contours (C.L) for the phantom scalar field cosmological parameters. We present the results for the three baselines considered: Planck 2018+BAO+Pantheon+Cosmic clocks, SPT-3G+WMAP9+BAO+Pantheon+Cosmic clocks, and ACTPol DR-4+WMAP9+BAO+Pantheon+Cosmic clocks.\label{fig:ScalarPhantomFieldconfidence}}
\end{figure}  

\begin{figure}[h]
\centering
\includegraphics[width=12.cm]{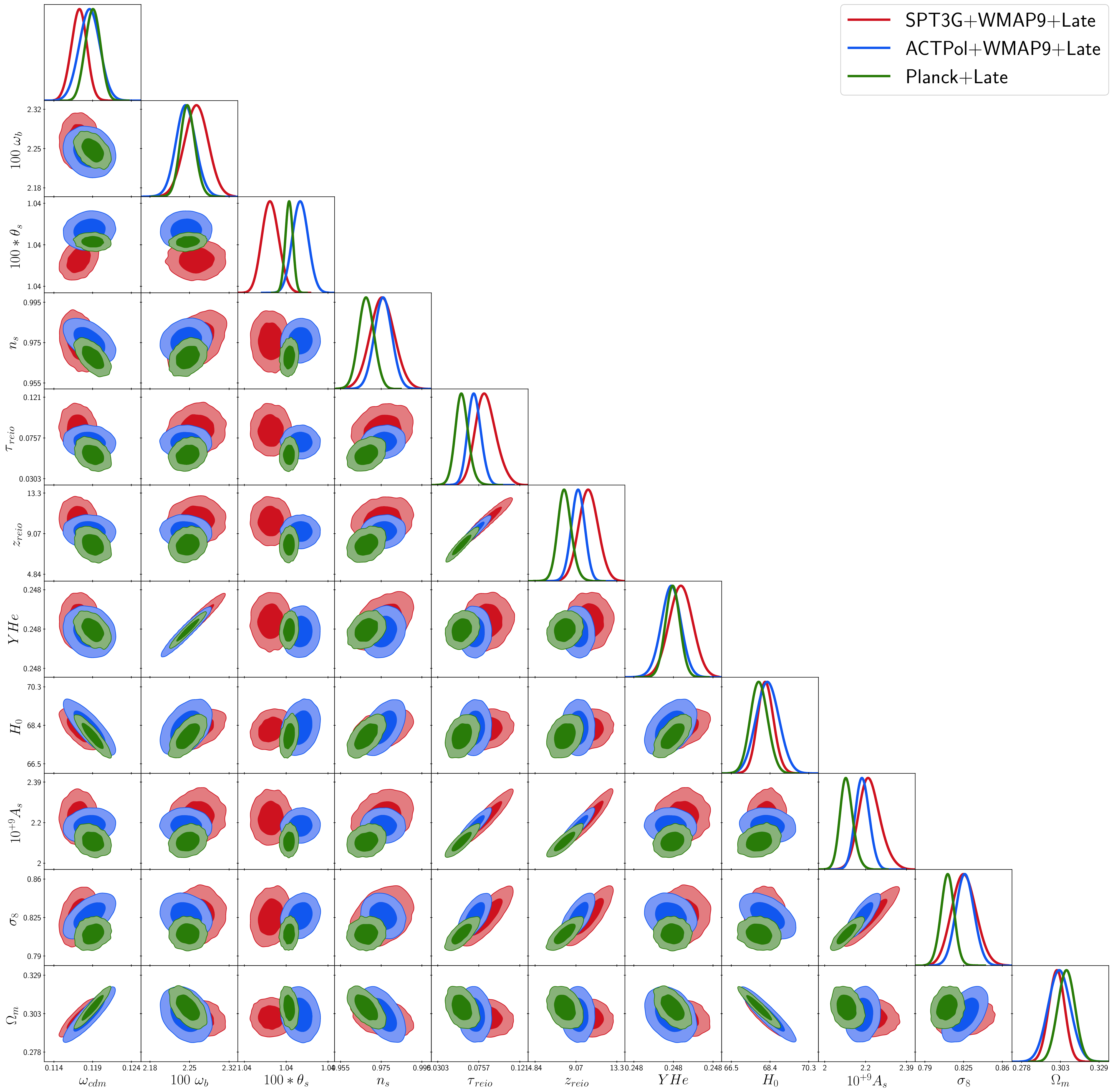}
\caption{$1 \sigma$ and $2\sigma $ confidence contours (C.L) for the $\Lambda$CDM cosmological parameters. We present the results for the same baselines as for the phantom scalar field.\label{fig:LCDMconfidence}}
\end{figure} 

\section{Cosmological tensions analysis}
\label{sec:results}

We ran Monte Carlo Markov Chain (MCMC) methods to constrain the cosmological parameters. Furthermore, we repeated this analysis for both the tracker scalar field and the $\Lambda$CDM model, to compare their performances. Moreover, we checked whether the tracker models reduced the $H_0$ tension compared to the last result from the SH0ES collaboration \cite{riess2022comprehensive}. For a statistical comparison between models, we computed the Hubble tension with
\begin{equation}
    T_{H_0} = \frac{|H_{0 \, \text{SH0ES}} - H_{0 \, \text{model}}|}{\sqrt{\sigma^2_{H_0 \, \text{SH0ES}}+\sigma^2_{H_0 \ \text{model}}}}.
\end{equation}
We sampled the cosmological parameters using the Metropolis-Hastings method and used Gelman-Rubin's convergence criterion $R-1<0.03$ \cite{gelman1992inference}. 

We sampled over the free parameters $100\omega_b, \omega_{\text{cdm}}, 100\theta_s, \ln (10^{10} A_s), n_s, \tau_{\text{reio}}$, the free scalar field parameters $\alpha_0, \alpha_1, \alpha_2$, the nuisance parameters of the experiments considered and the derived parameters $z_{\text{reio}}, \text{YHe}, H_0, 10^9 A_s, 
\sigma_8$, $\Omega_m$. For the scalar field parameters, we consider uniform prior probabilities $\alpha_0 = [-12,12], \alpha_1 = [-8,8], \alpha_2 = [1,16]$. These ranges were also considered to study tracker scalar field models \cite{cedeno2021tracker}. Moreover, these priors also simplify the shooting \texttt{CLASS} method to compute the value of the derived parameter $\Omega_\phi$ \cite{cedeno2021tracker}, and they are also consistent with the expected values of the potentials reported in \cite{roy2018new}. 

We present the mean values along with the $1\sigma$ C.L. uncertainties of the cosmological parameters for the phantom scalar field model and $\Lambda$CDM model in Tables \ref{tab:cosmologicalParametersPhantomScalarField} and \ref{tab:cosmologicalParametersLCDM}, respectively. We reported the results for the three baselines that we are considering: Planck 2018+BAO+Pantheon+Cosmic clocks, STP-3G+WMAP9+BAO+Pantheon+Cosmic clocks and ACTPol DR-4+WMAP9+BAO+Pantheon+Cosmic clocks. We also computed the tension between the Hubble constant derived from the MCMC and the latest SH0ES value of $H_0$ \cite{riess2022comprehensive}. In Figures \ref{fig:ScalarPhantomFieldconfidence} and \ref{fig:LCDMconfidence} we show the confidence contours at $1\sigma$ and $2\sigma$ for the cosmological parameters and the phantom scalar field and $\Lambda$CDM, respectively. 

As we can see, the three phantom scalar field parameters did not converge well for the compressed Planck 2028 baseline. However, the confidence contours show that they tend to have high values. For the remaining baselines, the parameter $\alpha_2$ is well constrained. However, the other two have high values as in the case of the Planck baseline. The rest of the cosmological parameters show good convergence. From the CMB catalogues considered in this work, Planck has the heavier baseline. Thus, we expect its confidence contours to be the smallest. We can see in Figures \ref{fig:ScalarPhantomFieldconfidence} and \ref{fig:LCDMconfidence} that this is indeed the case. 

When considering the phantom scalar field, the Hubble tension gets reduced for the three baselines. However, for the Planck+Late baseline, it is still higher than $4\sigma$ C.L. Note that this is different from the result from \cite{cedeno2021tracker} where they obtained $T_{H_0} = 3.63$ using a compressed Planck likelihood. However, in our results, by using the full Planck likelihood alongside late-time data, we notice that these phantom dark energy models do not address the Hubble tension. However, the tension was reduced considerably for the remaining baselines that did not include Planck. Furthermore, it is interesting that the tension got below the $3\sigma$ C.L. This shows that tracker phantom scalar field dark energy models can reduce the Hubble constant tension for CMB baselines that do not include the Planck likelihood. For these kinds of models, $w < -1$, which can produce a larger late-time expansion of the universe and thus, a higher value of $H_0$. However, this particular model cannot fully solve the problem since its tension is higher than $2\sigma$ C.L. 

\begin{figure}
    \centering
    \includegraphics[scale=0.8]{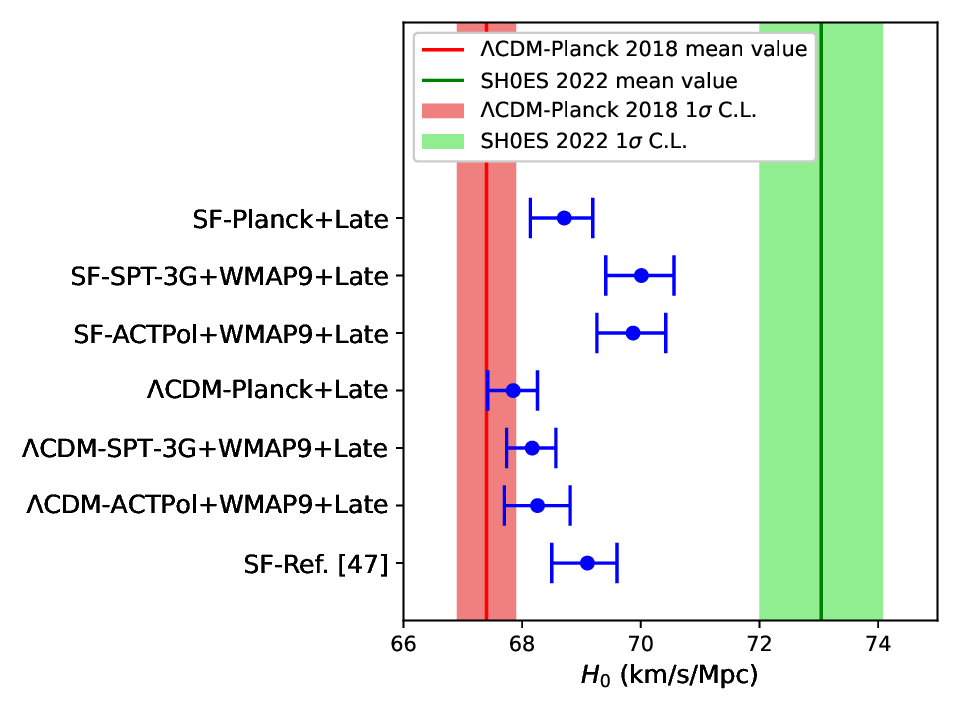}
    \caption{Whisker plot with the mean values and uncertainties at $1\sigma$ C.L. for the phantom scalar field (abbreviated as SF) and the $\Lambda$CDM models and the three baselines described. We also included the mean value and uncertainty at $1\sigma$ C.L. from the latest results from SH0ES collaboration \cite{riess2022comprehensive} and Planck 2018 \cite{aghanim2020planck} to study the Hubble tension. A result from \cite{cedeno2021tracker} was added for comparison. }
    \label{fig:whisker-plot}
\end{figure}

To visualize these results, we present Figure \ref{fig:whisker-plot}. We included the results of the Hubble constant for the phantom scalar field (abbreviated SF) and $\Lambda$CDM for the three baselines considered. We also included the result from \cite{cedeno2021tracker} of this model by using a compressed Planck likelihood. Finally, we include the mean values and uncertainties at $1\sigma$ for Planck 2018 \cite{aghanim2020planck} and SH0ES 2022 \cite{riess2022comprehensive}. As we can see, $\Lambda$CDM is consistent with Planck 2018 for the three baselines and it is in tension with SH0ES 2022. On the other hand, the phantom scalar field gets a higher value of $H_0$ that departs from the $\Lambda$CDM value but does not get high enough to be fully consistent with SH0ES 2022. However, other phantom scalar field models might be consistent with a higher $H_0$ by choosing a different parameterisation of $y_2$ or a different potential. 

\section{Conclusions}
\label{sec:final}

In this work, we studied new early constraints on tracker phantom scalar field cosmologies, in particular, some solutions regarding the proposal discussed in \cite{cedeno2021tracker}. We focused on the case where the late-time accelerated expansion of the universe is solely caused by a scalar field without the assistance of a cosmological constant. We employed early-time catalogues regarding CMB measurements and a full late-time catalog which includes SNIa, cosmic clocks and BAO observables. For these models, the EoS of the scalar field gives $w<-1$, which enables a larger cosmic late-time expansion and then a possibly higher value of $H_0$. 

We considered three model-independent CMB catalogues: WMAP9, SPT-3G and ACTPol DR-4, alongside late-time catalogues. Our methodology in this part consisted of constraining the tracker phantom scalar field using three different baselines: Planck+Late, STP-3G+WMAP9+Late, and ACTPol+WMAP9+Late. To compare our analysis, we computed the constraints with $\Lambda$CDM as a baseline model. The free parameters of the scalar field, named $\alpha_0, \alpha_1, \alpha_2$ from Eq.~\ref{eqn:y_2} did not fully converge for the Planck 2018 baseline. For the remaining baselines, $\alpha_0, \alpha_1$ did not obtain a full convergence. However, they show a good performance for the case of $\alpha_2$. The remaining cosmological parameters of the model showed good convergence for all the baselines and both the scalar field and $\Lambda$CDM models.

We were particularly interested in the constraints on the Hubble constant $H_0$ to study its statistical tension. For the Planck 2018 catalogue, the scalar field model shows a tension higher than $4\sigma$ C.L. when compared with the latest result from the SH0ES collaboration \cite{riess2022comprehensive}. This result differs from the one reported in \cite{cedeno2021tracker}, where a study was made using a compressed Planck likelihood. However, for the remaining baselines (SPT-3G+WMAP9+Late and ACTPol+WMAP9+Late), the $H_0$ tension was reduced under $3\sigma$ C.L, without altering the $\sigma_8$ value. Our results show that phantom scalar field models can reduce the $H_0$ constant tension when considering CMB model-independent baselines. However, it is important to mention that this particular model cannot completely solve the tension issue since the tension is not lower than $2\sigma$. We expect that new definitions in (\ref{eq:sys}), could lower the statistical tension at early times, e.g. from the ones reported in \cite{Cai:2009zp}. This aspect will be reported elsewhere.


\begin{acknowledgments}
This research has been carried out using computational facilities procured through the Cosmostatistics National Group ICN UNAM project. CE-R acknowledges the Royal Astronomical Society as FRAS 10147. 
JAN acknowledges financial support from the ``Excellence Project
Scholarship'' funded by the Physics and Astronomy Department of the University of Padova. 
This article is based upon work from COST Action CA21136 Addressing observational tensions in cosmology with systematics and fundamental physics (CosmoVerse) supported by COST (European Cooperation in Science and Technology). 

\end{acknowledgments}


\bibliography{apssamp}

\end{document}